\input harvmac.tex
\input amssym.tex

\def\newdate{Feb 2006}

\def\a{\alpha}
\def\b{\beta}
\def\g{\gamma}
\def\l{\lambda}

\def\p{\partial}

\def\CM {{\cal M}}

\def\p{\partial}

\font\cmss=cmss10
\font\cmsss=cmss10 at 7pt
\def\IL{\relax{\rm I\kern-.18em L}}
\def\IH{\relax{\rm I\kern-.18em H}}
\def\IR{\relax{\rm I\kern-.18em R}}
\def\inbar{\vrule height1.5ex width.4pt depth0pt}
\def\IC{\relax\hbox{$\inbar\kern-.3em{\rm C}$}}
\def\rlx{\relax\leavevmode}

\def\ZZ{\rlx\leavevmode\ifmmode\mathchoice{\hbox{\cmss Z\kern-.4em Z}}
 {\hbox{\cmss Z\kern-.4em Z}}{\lower.9pt\hbox{\cmsss Z\kern-.36em Z}}
 {\lower1.2pt\hbox{\cmsss Z\kern-.36em Z}}\else{\cmss Z\kern-.4em
 Z}\fi} 
\def\IZ{\relax\ifmmode\mathchoice
{\hbox{\cmss Z\kern-.4em Z}}{\hbox{\cmss Z\kern-.4em Z}}
{\lower.9pt\hbox{\cmsss Z\kern-.4em Z}}
{\lower1.2pt\hbox{\cmsss Z\kern-.4em Z}}\else{\cmss Z\kern-.4em
Z}\fi}

\def\CM {{\cal M}}

\def\CM {{\cal M}}

\font\manual=manfnt 
\def\dbend{\lower3.5pt\hbox{\manual\char127}}

\def\IZ{\relax\ifmmode\mathchoice
{\hbox{\cmss Z\kern-.4em Z}}{\hbox{\cmss Z\kern-.4em Z}}
{\lower.9pt\hbox{\cmsss Z\kern-.4em Z}} {\lower1.2pt\hbox{\cmsss
Z\kern-.4em Z}}\else{\cmss Z\kern-.4em Z}\fi}


\def\p{\partial}

\def\bar{\overline}

\def\rt2{\sqrt{2}}
\def\irt2{{1\over\sqrt{2}}}


\lref\GrassiTZ{
P.~A.~Grassi, G.~Policastro and P.~van Nieuwenhuizen,
``The massless spectrum of covariant superstrings,''
JHEP {\bf 0211}, 001 (2002)
[arXiv:hep-th/0202123];
}

\lref\GrassiUG{
P.~A.~Grassi, G.~Policastro, M.~Porrati and P.~van Nieuwenhuizen,
``Covariant quantization of superstrings without pure spinor constraints,''
JHEP {\bf 0210}, 054 (2002)
[arXiv:hep-th/0112162];
}

\lref\GrassiXV{
P.~A.~Grassi, G.~Policastro and P.~van Nieuwenhuizen,
``On the BRST cohomology of superstrings with / without pure spinors,''
Adv.\ Theor.\ Math.\ Phys.\  {\bf 7}, 499 (2003)
[arXiv:hep-th/0206216].
}

\lref\BerkovitsFE{ 
N.~Berkovits, 
``Super-Poincar\'e covariant quantization of the superstring'', 
JHEP { 0004}, 018 (2000) 
[hep-th/0001035]. 
}

\lref\WittenPX{
  E.~Witten,
  ``Two-dimensional models with (0,2) supersymmetry: Perturbative aspects,''
  arXiv:hep-th/0504078.
}
 
\lref\NekrasovWG{
  N.~A.~Nekrasov,
  ``Lectures on curved beta-gamma systems, pure spinors, and anomalies,''
  arXiv:hep-th/0511008.
} 

\lref\BerkovitsHY{
  N.~Berkovits and N.~A.~Nekrasov,
  ``The character of pure spinors,''
  Lett.\ Math.\ Phys.\  {\bf 74}, 75 (2005)
  [arXiv:hep-th/0503075].
}

\lref\BerkovitsBW{
  N.~Berkovits and S.~A.~Cherkis,
  ``Pure spinors are higher-dimensional twistors,''
  JHEP {\bf 0412}, 049 (2004)
  [arXiv:hep-th/0409243].
}

\lref\GrassiJZ{
  P.~A.~Grassi and J.~F.~Morales,
  ``Partition functions of pure spinors,''
  arXiv:hep-th/0510215.
}

\lref\GrassiSB{
  P.~A.~Grassi and N.~Wyllard,
  ``Lower-dimensional pure-spinor superstrings,''
  JHEP {\bf 0512}, 007 (2005)
  [arXiv:hep-th/0509140].
}

\lref\ChestermanEY{
  M.~Chesterman,
  ``Ghost constraints and the covariant quantization of the superparticle  in
  ten dimensions,''
  JHEP {\bf 0402}, 011 (2004)
  [arXiv:hep-th/0212261].
}

\lref\KoszulAB{
J.-L.~Koszul 
``Homologie et cohomologie des alg\'ebres de Lie''. 
Bulletin de la Soci\'et\'e Math\'ematique de France, 78 (1950), p. 65-127 
}

\lref\FischDQ{
  J.~M.~L.~Fisch, M.~Henneaux, J.~Stasheff and C.~Teitelboim,
  ``Existence, Uniqueness And Cohomology Of The Classical BRST Charge With
  Ghosts Of Ghosts,''
  Commun.\ Math.\ Phys.\  {\bf 120}, 379 (1989).
}

\lref\NekrasovVW{
  N.~Nekrasov and S.~Shadchin,
  ``ABCD of instantons,''
  Commun.\ Math.\ Phys.\  {\bf 252}, 359 (2004)
  [arXiv:hep-th/0404225].
}

\lref\SchwarzAK{
  A.~Schwarz,
  ``Sigma models having supermanifolds as target spaces,''
  Lett.\ Math.\ Phys.\  {\bf 38}, 91 (1996)
  [arXiv:hep-th/9506070].
}

\lref\WittenNN{
  E.~Witten,
  ``Perturbative gauge theory as a string theory in twistor space,''
  Commun.\ Math.\ Phys.\  {\bf 252}, 189 (2004)
  [arXiv:hep-th/0312171].
}

\lref\AlekseevNP{
  A.~Alekseev and T.~Strobl,
  ``Current algebra and differential geometry,''
  JHEP {\bf 0503}, 035 (2005)
  [arXiv:hep-th/0410183].
}

\lref\BerkovitsPX{
  N.~Berkovits,
  ``Multiloop amplitudes and vanishing theorems using the pure spinor
  formalism for the superstring,''
  JHEP {\bf 0409}, 047 (2004)
  [arXiv:hep-th/0406055].
}

\lref\WyllardFH{
  N.~Wyllard,
  ``Pure-spinor superstrings in d = 2, 4, 6,''
  JHEP {\bf 0511}, 009 (2005)
  [arXiv:hep-th/0509165].
}
\lref\KapustinPT{
  A.~Kapustin,
  arXiv:hep-th/0504074.
}

{\Title{
\vbox{\hbox{CERN-PH-TH/2006-020}
\hbox{LMU-ASC 07/06}
\hbox{DISTA-2006}
}}   
{\vbox{
\centerline{Curved Beta-Gamma Systems}
\smallskip
\centerline{and Quantum Koszul Resolution}
}}  
 
\medskip\centerline
{
P.~A.~Grassi$^{~a,b,c,}$\foot{pgrassi@cern.ch} and G.~Policastro$^{~d,}$\foot{policast@theorie.physik.uni-muenchen.de}
} 

\bigskip   
\centerline{
$^{(a)}$ 
{\it Centro Studi e Ricerche E. Fermi, 
Compendio Viminale, I-00184, Roma, Italy,}}
\centerline{
$^{(b)}$ 
{\it CERN, Theory Unit, Physics Department, 
1211-CH Geneva 23, Switzerland,}}
\centerline{
$^{(c)}$ 
{\it DISTA,
Universit\`a del Piemonte Orientale,
via Bellini 25/g, I-15100, Alessandria, and}}
\centerline{\it
INFN - Torino, Gruppo Collegato di Alessandria, Italy.}}
\centerline{
$^{(d)}$ \it
Arnold Sommerfeld Center for Theoretical Physics}
\centerline{\it 
Ludwig-Maximilians-Universit\"at, 
Theresienstra{\ss}e 37, 80333 M\"unchen, Germany}

\bigskip
\bigskip

\noindent
We consider the partition function of beta-gamma systems in curved space of the type discussed by Nekrasov and Witten.
We show how the Koszul resolution theorem can be applied to the computation of the partition 
functions and to characters of these systems and find a prescription to 
enforce the hypotheses of the theorem at the path integral level. 
We illustrate the technique in a few examples: a simple 2-dimensional target space, the N-dimensional conifold, and a superconifold. Our method can also be applied to the Pure Spinor constraints of superstrings. 

\Date{\newdate}

\newsec{Introduction}

One of the major revolutions during the last years in string theory 
is certainly the advent of Pure Spinor formalism for Superstrings 
\BerkovitsFE. 
This formalism provides an alternative to the RNS and GS formulations 
and it has survived several checks of consistency. It 
is based on a set of variables $x^{m}, \theta^{\a}, p_{\a}$ describing the 
supertarget space (with a free action if the space is flat) 
 and a set of ghost fields $w_{\a}$ and $\l^{\a}$ which 
are constrained to leave on a subvariety defined by the so-called Pure Spinor constraints. The need of these constraints comes from nilpotency of the BRST 
charge $Q = \oint dz \, \l^{\a} (p_{\a} + \dots)$. 
The ghost fields are complex fields, but the action and the constraints are 
holomorphic and this fact allows us to use the recent analysis of 
\WittenPX\ and \NekrasovWG\ (see also \KapustinPT ) of curved beta-gamma systems 
and the chiral-de Rham complex. 

The ghost fields form a beta-gamma system whose action does not depend on a choice of metric. Nevertheless, they are not free fields because of the constraints. 
We are interested in computing the partition function of the system (it is always understood that states are weighted with a sign according to their parity, so we are really computing the character, often with additional twisting by other currents as well). 
There are several techniques available for this computation: the direct counting of the states, the computation of the cohomology, or localization with respect to group actions on the target space.  

One of the authors, in collaboration with J.F. Morales, has computed in 
\GrassiJZ\ the 
the zero-modes and the non-zero modes sector (first level) of the  
character of pure spinor beta-gamma systems 
(for the 10d superstrings and for lower dimensional systems \GrassiSB). 
For the latter the counting of the states is used to compute the 
character and this technique is simpler than the brute-force cohomology computations. Unfortunately, even this technique does not permit the 
full computation of the partition function of the pure spinor beta-gamma system 
since one should be able to derive a closed formula for 
the dimensions of each representations and then sum these 
contributions with appropriate weight. 

In order to overcome these problems, one may try to use the Koszul theorem 
to resolve the constraint adding new ghost-for-ghosts to the theory 
\refs{\KoszulAB,\FischDQ}. This 
program was started for the superstrings in papers \refs{\GrassiUG,\GrassiTZ,\GrassiXV} and it was followed by 
\refs{\ChestermanEY,\BerkovitsHY} where an infinite set of new fields has been proposed.
At first sight this may appear a satisfactory solution, but a closer analysis quickly reveals a problem, namely that the correct 
cohomology is recovered only by suitably restricting the functional space of fields (crucially, including the conjugate momenta) with the help of a new quantum number denoted {\it grading}. One can easily show that without this restriction the cohomology turns out to be wrong, in fact often it is empty. 

In this context it is perhaps useful to note that as long as one does not include the contributions of operators constructed using the conjugate momenta, the partition function has a simple geometric interpretation: at the massless level it is the holomorphic Euler character of the variety defined by the constraints (the equivariant character when there is a twisting by currents generating an action on the target space), and the higher modes give the equivariant character of the loop space of the same variety. In the latter case, the group action includes an $S^{1}$ factor given by the rotations of the loop. However the inclusion of the conjugate momenta spoils the geometric interpretation. This fact has been recognized in \NekrasovWG\ and \AlekseevNP.

This problem does not seem to pertain only to the pure spinor 
string theory, but it is related to the implementation of the Koszul resolution 
technique to a quantum system such as a generic beta-gamma system. 
The case of the pure spinor is a particularly interesting application, but it is a rather complicated case, as pure spinors are defined by a set of constraints which is not irreducible (in other words, the variety is not a complete intersection). 
We consider in this note some simpler beta-gamma systems where the variety is a hypersurface, so it is defined by a single 
constraint. This is already sufficient to see the problems we are trying to solve. 
In particular we consider the example of the conifold. Even though this is a singular variety, the singularity will not present a problem for our purposes. 

We compute the spectrum of some simple beta-gamma systems 
by constructing the vertex operators 
(or the corresponding states) compatibly with the contraints. This amounts to 
listing all possible states modulo the constraints. At the level of zero modes, 
this can be easily done by using the localization techniques \BerkovitsHY, 
but it cannot be implemented for the non-zero modes. 

We found that by keeping track of the grading number in the partition function, and removing all inverse powers of the new grading number, the 
resulting partition function yields the correct expression. This prescription  
can be easily implemented at the level of partition function constructed 
in terms of free fields. In order to check the formula, we compute 
the spectrum level-by-level  in three different models and we show that 
the character of these multiplets coincide. In sec. 2, we give the basis and 
the formula for the character computation. In sec. 3, we analyze three examples. 
In sec. 4, we add some considerations and outlook.  

\newsec{Action and Constraints}

\subsec{Curved beta-gamma systems}

We briefly review some ingredients of the formulation of a beta-gamma 
system on a manifold ${\cal M}$. We assume that the manifold is parametrized by a set 
coordinates $\l^{i}$ ($i= 1, \dots, d$ where $d$ is the dimension of the manifold) and we assume for the moment that the manifold is flat. 
We add a set of fields $w_{i}$, conjugated to $\l^{i}$, 
which are worldsheet one-forms 
$w_{i} = w_{i \mu} dx^{\mu}$ and we postulate  the free action  
\eqn\parAB{
S =  {1\over 2 \pi} \int d^{2}z \, w_{i} \bar \p\l^{i}\,,
}
in the conformal gauge (we use the complex coordinates 
$z, \bar z$ for the worldsheet). The beta-gamma system differs from the more usual sigma-model in that the action is holomorphic and depends only on the field $\lambda^{i}$ and $w_{i}$ and not on their complex conjugates. Therefore it can be defined on any manifold with a complex structure, without the need for a metric or a K\"ahler structure. 
The free OPE is simply given by 
\eqn\freeOPE{\l^{i}(z) w_{j}(w) \sim 
 (z-w)^{-1} \delta^{i}_{j}.}
The system is a conformal field theory with 
central charge $c = +2 d$ and it has a global symmetry generated by $J^{w\l} = :\l^{i} w_{i}:$ which assigns charge $+1$ to $\lambda^{i}$ and 
$-1$ to $w_{i}$. It is convenient to give also the mode expansion of the fields 
\eqn\modeA{
\l^{i}(z)= \sum_{n} \l^{i}_{n} z^{-n}\,, ~~~~~
w_{i}(z) = \sum_{n} w_{i, n} z^{-n-1}\,,
}
and the vacuum is chosen by $\l^{i}_{-n} |0\rangle =0, \forall n > 0$ and 
$w_{-n,i} |0\rangle =0, \forall n \geq 0$. We denote by $\hat M$ the Fock space. 

The partition function (character) of this system is defined by 
$Z(q|t) = {\rm Tr}_{\widehat\CM}( q^{L_{0}} t^{J^{w\l}_{0}} )$ where $J^{w\l}_{0} = 
\oint dz \l^{i} w_{i}$ and $L_{0} = \oint dz z T_{zz}$ with $T_{zz} = w_{i z}
\p_{z} \l^{i}$. Here $q$ is the modular parameter and $t$ is the parameter 
corresponding to the $U(1)$ current $J^{w\l}$. $Z$ can be readily computed  for the free 
system:
\eqn\parA{
Z(q|t)^{-1} = (1-t)^{d} \prod_{n=1}^{\infty}(1 - q^{n} t)^{d} (1 - q^{n} t^{-1})^{d}  = 
{
\vartheta_{1}(\nu|\tau)^{d} e^{-{i \pi \tau d\over 6}} 
\over 
\eta(\tau)^{d} e^{ - i \pi (\nu + {1\over 2})d} 
}   \,.
}
where $q = \exp(2 i \pi \tau)$ and $t = \exp(2 i \pi \nu)$.   
We now pass to a curved target space. In this case, the fields $w_{i},\l^{i}$ form a non-linear 
sigma model that can be affected by anomalies, as shown by Witten \WittenPX. We recall the points of his analysis that are needed for the present purposes. 
The model can be analyzed locally by covering the manifold with a set of charts. In each chart $U_{\a}$ there is a free CFT described by the action \parAB with coordinate and momenta $\l_{i \a},w_{i \a}$.  These local theories must be glued together across the different charts to get a globally defined theory. 
The coordinates $\l^{i}$ have classical transformation laws: $\l^{i}_{\a} = f^{i}_{\a\b}(\l^{j}_{\b})$ where $f_{\a\b}$ are the holomorphic transition functions defined on $U_{\a}\cap U_{\b}$.
The conjugate momenta are classically (1,0)-forms on the target space and as such they should transform with the jacobian of the transition function matrix: 
$$ w_{i,\a} = {\partial \l^{j}_{\b} \over \partial \l^{i}_{\a}} w_{j,\b} \, .
$$
In general, this transformation rule is not compatible with the requirement that both $w_{i \a},\l_{i \a}$ and $w_{i \b}, \l_{i \b}$ are conjugate pairs that satisfy the free OPE \freeOPE. The reason is that 
$w_{i \b}$ is defined in terms of $w_{j \a},\l_{j \a}$ as a composite operator that requires a regularization, and this has to be done covariantly. One must look for a correction to the 
transformation law: 
\eqn\transfw{
w_{i,\a} = {\partial \l^{j}_{\b} \over \partial \l^{i}_{\a}} w_{j,\b} + B_{\a\b,ij} \partial \l^{j}_{\beta} \,.
}
The problem of finding a set of $B_{\a\b,ij}$ consistently can be recast in a problem of \v Cech cohomology, and the obstruction is parametrized by $H^{2}({\cal M},\Omega^{2})$ (see \WittenPX, \NekrasovWG). In fact, it is given by the Pontriagin class $p_{1}({\cal M})$. There is also an anomaly proportional to $c_{1}({\cal M}) c_{1}(\Sigma)$ where $\Sigma$ is the worldsheet. The examples we will consider are anomaly-free, nevertheless our method will allow us to work always in global coordinates. 

Let us now specialize to the case where the manifold ${\cal M}$ is a hypersurface in $V \approx {\Bbb C}^{N}$, so it is defined by an algebraic constraint for the fields $\l^{i}$  of the form 
$\Phi(\l) =0$. The constraint is non-degenerate if $\p_{\lambda^{i}} \Phi(\l)|_{\Phi(\l) =0} \neq 0$ and it is not reducible if there is no function $F(\l)$ such that $F(\l) \Phi(\l) =0$ for $\Phi(\l) \neq 0$. 
The constraint $\Phi(\l)$ is a first class constraint and generates the 
gauge symmetry (with local parameter $\Lambda(z)$)
\eqn\parC{
\delta_{\Lambda} w_{i} = \Lambda(z) {\partial\over \partial \lambda^{i}} \Phi(\l) \,,
}
For example, we can consider the constraint 
\eqn\parCA{
\Phi(\l) = \l^{i} g_{ij} \l^{j} = 0\,,
}
 and therefore the gauge symmetry becomes $\delta_{\Lambda} w_{i} = \Lambda g_{ij} \lambda^{j}$ where $g_{ij}$ is a metric on the target 
 space. For $i=1,\dots,4$ and for the euclidean metric $g_{ij}=\delta_{ij}$, the 
 constraint \parCA\ is the conifold which is an hypersurface in ${\bf C}^{4}$. 
 The action and the constraints are holomorphic and therefore we can apply the analysis on Witten  \WittenPX\ and Nekrasov \NekrasovWG. Notice that 
 if instead of a single holomorphic constraint we consider a set of them 
 $\l^{i} g^{\Sigma}_{ij} \l^{j} = 0$ (where $\Sigma$ denotes the set of constraints) 
 we reproduce the pure spinor constraints of superstrings \BerkovitsFE.\foot{The pure spinor constraints of superstrings are given in terms of a  Weyl spinor $\l^{\a}$ of $SO(1,9)$  and they read $\l^{\a} \g^m_{\a\b} \l^{\b} =0$. They 
 are holomorphic since they depend only upon $\l^{\a}$ and they are 
 reducible $A^{\a}\g^{m}_{\a\b} \l^{\b} (\l \g_{m} \l) =0$ for any $A^{\a}$  
 because of the Fierz identities.  Lower dimensional pure spinors 
 are studied in \refs{\GrassiJZ,\GrassiSB,\BerkovitsHY,\BerkovitsBW}. 
 }  
  
The constraint \parCA\ reduces the number of independent coordinates 
$\lambda^{i}$, and one should locally choose a parametrization and construct a solution to \parCA\ in each patch in terms of independent coordinates and 
then show that the observables can be extended to all charts and that there are no anomalies \refs{\WittenPX,\NekrasovWG}, in keeping with the general philosophy. Of course the space has a singularity that will not be covered by any chart. In order to 
construct the spectrum of the theory, one can then proceed in two ways: {\it i)} 
covariantly, without solving \parC, but implementing it on the Fock 
space generated by the modes $\l_{n}^{i}$ and $w_{n,i}$ (see 
also \GrassiJZ\ for a similar counting) and {\it ii)} by solving the constraint 
and working on a single patch. However, the second way has to be supplemented by a condition to sew the states given on a patch and those on an another one. 

If we proceed covariantly, we must take into account the gauge invariance \parC. 
It is convenient to define  $\l_{i} = g_{ij} \l^{j}$.\foot{To compute the 
OPE's in the case of constraints, it is convenient to 
chose a parametrization. Decomposing $\l^{i} = (\l^{0}, \l^{I})$ with 
$I=1, \dots, N-1$, the constraint \parCA\ is solved by $\l^{0} =   
\sqrt{-(\l^{I})^{2}}$. Using the gauge symmetry \transfw\ we set $w_{0}=0$. 
The OPE becomes $w_{i}(z)  \l^{j}(w) \sim - (z-w)^{-1} \Big(\delta_{i}^{I} \delta_{I}^{j} - {1\over \sqrt{-(\l^{I})^{2}}} 
\l_{I} \delta^{I}_{i} \delta^{j}_{0} \Big)$ where the second term 
is needed to enforce the constraint and it is necessarily non-covariant. 
The complete conformal field theory analysis will be given elsewhere.
} 
There are two types of gauge invariant combinations 
\eqn\padCB{
J_{[ij]} = :w_{[i} g_{j]k}\lambda^{k}:\,, ~~~~~
J = :w_{i} \lambda^{i}:\,}
where $A_{[i}B_{j]} = {1\over 2} ( A_{i}B_{j} - A_{j}B_{i})$.  The currents 
$J_{ij}$ generate the $SO(N)$ rotations. 
Notice that these two operators 
are not independent. In fact, at the classical level, $J_{[ij]} \l^{j} + {1\over 2} \, J \lambda_{i} = 0$. However, at the quantum level this Ward identity is deformed into 
\eqn\parD{
:J_{[ij]} \l^{j}: + {1\over 2} :J \lambda_{i}: = {2-N\over 2} \p \l_{i}\,,   
}
As a consequence, the stress energy tensor $w_{i} \p \l^{i}$ (which is 
gauge invariant on the surface of the constraints $\Phi(\l)=0$)
can be rewritten as \refs{\BerkovitsPX,\WyllardFH}
\eqn\parE{
T_{w,\l} = w_{i} \p\l^{i} = -  
:J_{[ij]} J^{[ij]}: + {1\over 2} :J^{2}: + {2 -N\over 2} \p J\,,
}
where the coefficients are fixed by computing the conformal weights 
of the fields according to their representation under the 
transformations generated by $J_{[ij]}$ and under the $U(1)$ charge. 
The last term comes from the normal ordering. In the same 
way, any gauge invariant operator can be expressed in terms of 
$J_{[ij]}$ and $J$. This fact, toghether with the observation 
that there is only the field $w_{i}$ which carries the negative $U(1)$ charge, implies that there is no vertex operator/state with negative charge. 

\subsec{The Koszul-Tate resolution}

The Koszul-Tate resolution provides a way to implement the constraint 
$\Phi(\l)=0$ through its homology \KoszulAB. 
That is,  one has to find a complex with a differential $\delta$ such that 
$H_{0}(\delta)$ is equal to the algebra of holomorphic functions 
${\cal H}ol({\cal M})$ and $H_{k}(\delta) =0$ for $k\neq 0$. 
${\cal H}ol({\cal M})$ can be identified with the quotient algebra ${\cal H}ol({V})/ {\cal N}$ of the space 
of holomorphic functions on the vector space ${V}$ 
that differ by an element of the ideal ${\cal N}$ of functions vanishing on ${\cal M}$. So, to fullfill $H_{0}(\delta) \equiv 
(Ker \, \delta)_{0}/(Im \, \delta)_{0} =  {\cal H}ol({V})/ {\cal N}$ it is 
natural to define $\delta$ so that $(Ker \, \delta)_{0} = {\cal H}ol({V})$ and 
$(Im \, \delta)_{0} = {\cal N}$. 

To achieve the first condition, 
one simply sets $\delta \l^{i} =0$ and assigns them grading zero. 
For the second one, one introduces a new worldsheet 
zero-form $c$ for each constraint and sets 
$\delta c = \Phi(\l)$. Therefore, any function $F(\lambda)$ vanishing on the constraint can be written as $F (\l) = f(\l) \Phi(\l) = \delta ( f(\l) c)$ and 
$F(\l) \in (Im \, \delta)_{0}$. To satisfy the grading properties on $\delta$, one 
assigns charge  $+1$ to $c$ and extends $\delta$ to arbitrary polynomials 
on ${\cal H}ol(V) \otimes {\bf C}[c]$ by requiring that $\delta$ be an odd derivation. The differential $\delta$ is known as Koszul-Tate differential. 
One of the main hypothesis of the Koszul-Tate theorem is the 
absence of negatively charged states. However, for a quantum system, there might be the possibility to construct negatively charged states (such as 
the conjugate momenta to $c$) and that would violate the theorem. We discuss this issue presently.  First we  translate the Koszul-Tate differential into a 
charge of the conformal field theory, then we analyze its cohomology.  

In terms of beta-gamma system, the $\delta$ is represented by 
an anticommuting charge $\widehat\delta = \oint dz \, J^{K}_{z}$ constructed from a new current
$J^{K}_{z} = b_{z} \Phi(\l)$ where $b_{z}$ is the conjugate momentum 
of $c$. The new b-c system has conformal charge $-2$ 
and is described by the holomorphic free action 
\eqn\bc{
S_{bc} = {1\over 2\pi} \int d^{2}z \, b_{z} \p_{\bar z} c\,. 
}
The current $J^{K}_{z}$ is nilpotent, $J^{K}_{z}(z) J^{K}_{z}(w) \sim 0$,  
and vanishes on ${\cal M}$. As for the beta-gamma system, there is a 
$(1,0)$ current $J^{bc}_{z} = :b_{z} c:$ which is a global symmetry of the 
action \bc.
The Koszul-Tate differential $\widehat\delta$ 
has charge $-1$ with respect to $J^{bc}$. 
In the following, we use $J_{z} = J^{w\l}_{z} + J^{bc}_{z}$ 
for the ghost number and 
$J^{g}_{z} = J^{w\l}_{z} + 2 \, J^{bc}_{z}$ 
for the grading number (in the literature this 
is also known as the antifield number). Using these definitions, it is 
easy to show that the action of $\widehat{\delta}$ on the 
fields is 
\def\WD{\widehat{\delta}}

\eqn\acA{
\WD \l^{i} =0\,, ~~~~
\WD w_{i} = b_{z} \p_{\l^{i}} \Phi(\l)\,, ~~~~
\WD c = \Phi(\l)\,, ~~~~
\WD b_{z} = 0\,.
}
where the variation reproduces the gauge symmetry generated by $\Phi$. Since 
$b_{z}$ is inert under $\WD$, any function 
$F(b, \p b, \p^{2} b, \dots, \l^{i}, \p \l^{i}, \dots)$ is invariant under $\WD$. 
This means that there are new cohomology classes such as $b, \p b, \p b \l^{i}, \dots$ 
which have to be eliminated since they do not represent the original 
spectrum computed with the constraints. In order to remove these new cohomology classes, one should add new fields: for instance, to eliminate $b_{z}$ from the cohomology one 
can introduce a new variable $\rho_{z}$ 
such that $\WD \rho_{z} = b_{z}$. This amounts to adding a new term to $\WD$ 
as follows 
\eqn\acB{
\WD \rightarrow \WD + \oint dz \, \tau b_{z}\,,
} 
where $\tau$  is the conjugate of $\rho_{z}$ with zero conformal weight.  
However, the similarity transformation generated by $\exp(- \oint dz \rho_{z} \Phi)$ removes the first term of $\WD$ leaving a very simple differential, 
whose cohomology is represented by any function $F(\l^{i}, \p \l^{i}, \dots, w_{i}, \p w_{i}, \dots)$ which does not depend on  $\rho_{z}, \tau, b_{z}, c$. The constraint disappeared and the spectrum coincides with of  the free theory \parAB, with  \parA\ as partition function. This procedure then does not give the correct result since the cohomology is formed by all states of the free theory.  

There is however another way to remove the unwanted cohomology classes 
and to recover the correct spectrum. This is done by 
computing the new character 
\eqn\acC{
Z(q|s) = {\rm Tr}_{\widehat\CM'}( q^{L_{0}} s^{J^{g}_{0}}) 
}
where $J^{g}_{0} = \oint dz \, J^{g}_{z}$ and $\widehat\CM'$ is the 
Fock space $\widehat\CM$ tensored with the Fock space of the b-c system. 
The correct partition function is now defined by 
\eqn\acD{
Z(q|t) = \lim_{s \rightarrow t} Z(q|s)|_{\rm regular}\,, 
}
$$Z(q|s) = {(1 - s^{2}) \over (1- s)^{N} }  
\prod_{n=1}^{\infty} {(1 - q^{n} s^{2}) (1 - {q^{n} s^{-2}}) 
\over 
(1 - q^{n} s)^{N} (1 - {q^{n}  s^{-1}})^{N} 
}$$
where $ Z(q|s)|_{\rm regular} = \sum_{k\geq 0} s^{k} Z_{k}(q)$ and the 
poles in $s$ are removed. The power of $s$ in the numerator of \acD\ is due to the {\it quadratic} constraint $\Phi(\l)$.  
The power of $t$ counts the ghost number 
of different states. This is in agreement with the previous considerations on the absence of negatively charged states in the spectrum. 
Expanding $Z(q|t) = \sum_{l\geq0} q^{l} Z_{l}(t)$ and expanding 
the coefficients $Z_{l}(t)$ in powers of $t$, one can read the number 
of states at each level $l$ weighted with the ghost number $t$. Of course, this 
does not tell us which states actually occur and it gives only the difference between fermonic and bosonic states, nevertheless it gives important 
information on the spectrum. 

The technique can be used also for reducible constraints. If 
the single constraint $\Phi$ is replaced by a set of them $\{\Phi^{\Sigma}\}$, we have 
to introduce a multiplet of b-c system carrying the index $\Sigma$. This would not be enough if there are relations between constraints. Infact, one 
has to introduce a set of ghost-for-ghosts $(c^{\Sigma_{k}}, 
b_{z}^{\Sigma_{k}})$ 
where the index $k$ denotes the level and the label $\Sigma_{k}$ denotes 
the multiplet of fields. One assigns to each multiplet the ghost and antighost numbers $k, -k$ respectively,  
and the partition function can be 
obtained by  $Z(q|t) = \lim_{s \rightarrow t} Z(q|s)|_{\rm regular}\,,$ 
where  
\eqn\acE{
Z(q|s) = \prod_{k=1}^{R} \Big[
{(1 - s^{M_{k}})^{N_{k}} \over (1- s)^{N} }  
\prod_{n=1}^{\infty} {(1 - q^{n} s^{M_{k}})^{N_{k}} (1 - {q^{n} s^{-M_{k}}})^{N_{k}} 
\over 
(1 - q^{n} s)^{N} (1 - {q^{n} s^{-1}})^{N} 
}\Big]
}
with $N_{k}$ the dimension (this number can be negative by taking 
into account  the statistic $+|N_{k}|$ for fermions and $-|N_{k}|$ for bosons) of the multiplet of ghost-for-ghost at level $k$, 
$R$ the maximum level and $M_{k}$ is the degree of the constraint at each level $k$. In the case of infinitely reducible constraints, the 
formula \acE\ needs a regularization procedure in order to 
guarantee the convergence of the infinite product over $k$. At the present stage 
we cannot verify this assertion. 

Notice that in the restricted functional space of fields with positive grading the similarity transformation 
discussed after eq.~\acB\ is no longer performable and this eliminates the 
paradox. 

\newsec{The examples}

\subsec{The cone}

Let us now implement the formula \acD\ for a very simple model. 
This model  has been already discussed in \GrassiJZ, but, here, we 
will consider only the pure spinor part without the fermions and the 
target space coordinates. We let $N=2$ and we adopt the euclidean metric 
on ${\cal M}$. In terms of light-cone coordinates, ($\l,\bar\l$), 
the constraint becomes
\eqn\conoA{
\Phi(\l) = \l \bar \l =0\,.
}
Notice that $\l$ and $\bar \l$ are not complex conjugates, but two independent complex variables. 
Even though the solution of the constraint is almost trivial, describing two lines intersecting at the origin, 
it has already interesting consequences on the computation of the spectrum. 
For example, restricting to the sector of  zero modes $\l_{0}$ and $\bar \l_{0}$, 
the Hilbert space is spanned by
\eqn\conoB{
{\cal H}_{l=0} = \{ 1, \l_{0}^{n}, \bar\l^{n}_{0}\}\,, ~~~~~ \forall n > 0 \,.
}
The character formula for this space is computed very easily and it gives 
$$Z_{l=0}(t) = (1+t)/(1-t) = 
1 + 2\,t + 2\,t^2 + 2\,t^3 + 2\,t^4 + {\cal O}(t^5)\,.$$
Let us move to the first massive states. The computation is again very easy since 
one can take into account that the dependence upon $w$ and $\bar w$ (the conjugate fields of $\l$ and $\bar\l$) is only through the gauge 
invariant combinations $J = w \l$ and $\bar J = \bar w \bar \l$. 
This means that at level 1  we find 
the states 
\eqn\conoC{
{\cal H}_{l=1} = \{
\bar\l_{1} \bar\l_{0}^{n}, \l_{1} \l_{0}^{n}, \l_{1}\bar \l_{0}, w_{1} \l_{0}^{n+1}, 
\bar w_{1} \bar\l_{0}^{n+1}\}  ~~~~~ \forall n > 0 \,.
}
and the character is given by 
$$Z_{l=1}(t) = {2 t \over (1+t)} + t^{2} +  {2 \over (1+t)}  = 
2 + 4\,t + 5\,t^2 + 4\,t^3 + 4\,t^4 + 4\,t^5 + {\cal O}(t^6).$$ 
The same result can be obtained 
by using the formula \acD\ which in the present 
case becomes
\eqn\conoD{
Z(q|t) = \lim_{s\rightarrow t}
\Big[{(1 - s^{2}) \over (1- s)^{2} }  
\prod_{n=1}^{\infty} {(1 - q^{n} s^{2}) (1 - {q^{n} s^{-2}}) 
\over 
(1 - q^{n} s)^{2} (1 - {q^{n}  s^{-1}})^{2} 
}\Big|_{\rm regular}\Big]
}
and the computation of $Z_{l=0}(t) $ and $Z_{l=1}(t)$ can 
proceed as illustrated above. Using \conoD\ one 
can compute easily higher contributions to $Z_{l}$. For instance, 
one has 
$$
Z_{l=2} =5 + 10\,t + 12\,t^2 + 12\,t^3 + 10\,t^4 + {\cal O}(t^{5})
$$
which is again in agreement with the computation of the spectrum by 
listing all the states. 

Incidentally, we can use this example to illustrate the fact that the singularity of the space is of small importance for our computations. For instance, we could imagine smoothing the space by taking the deformation $\l \bar \l =\mu$. We can of course solve the constraint by $\bar \l = \mu/\l$. Topologically this space is nothing but a cylinder
and we can use the coordinate $\phi = \ln \l$. At the massless level we have then the tachyon-like operators $e^{i n \phi}$, spanning the same space as in \conoB. At level 1 we have states $\partial \phi e^{i n \phi}, \omega e^{i n \phi}$. The state $\l_{1} \bar \l_{0}$ corresponds to the current $\partial \phi$. We see then that nothing is gained or lost by resolving the singularity. 

\subsec{The conifold}

Let us move to a more interesting example. We assume that $N=4$ and 
the constraint $\Phi$ becomes 
\eqn\coA{
\Phi(\l)  = \sum_{i=1}^{4} \l^{i} \l^{i} =0
}
which is a holomorphic hypersurface in ${\bf C}^{4}$. This represents the 
conifold space which is a singular manifold. For the specific value $N=4$ 
the manifold is Calabi-Yau, but our analysis does not require such condition. 
This is probably due to the specific nature of the sigma model on the conifold, which 
is not a conventional one. 

The computation of character formula based on counting of 
states can be easily done at level zero by observing that one simply   
has to count the polynomials $\l^{(i_{1}} \dots \l^{i_{n})}$ and  
subtract all possible traces. But this amount to selecting 
the irreducible representations of $SO(N)$ wich are totally symmetric and traceless. 
Therefore, the $l=0$ states modulo the constraint \coA\ are in correspondence with the Young tableaux having  one row with $k$ boxes. The dimension of the
corresponding representation is given by the classical formula (for $N$ even) 
\eqn\coB{
{\rm Dim}_{N}(k) = \prod_{j=2}^{N/2} 
{(k + N/2 - 1)^2 - (N/2 - j)^2 \over (N/2 - 1)^2 - (N/2 - j)^2}\,.
}
The character formula is obtained by resumming the dimensions 
weighted with $t^{k}$, and one gets
\foot{The equations for $SO(2 n +1)$ i.e. for $N$ odd are different, but 
the result of resummation is the same.} 
\eqn\coC{
Z_{l=0}(t) = \sum_{k=0}^{\infty} t^{k} {\rm Dim}_{N}(k)  = 
{(1-t^{2}) \over (1-t)^{N}} 
\,.
}
In the case $N=4$, this formula can be obtained also by localization technique 
as illustrated in \NekrasovVW. Using new coordinates $X = \l^{1} + i \l^{2}, Y= \l^{1} - i \l^{2}, 
T = \l^{3} + i \l^{4}, Z = \l^{3} - i \l^{4}$, the constraint reads $XY + TZ = 0$, and the equation is invariant under a torus action with three different charges $q_{1}, q_{2}, q_{3}$: 
\eqn\coD{
X \rightarrow q_{1} X\,, 
Y \rightarrow q_{2}q_{3} Y\,, 
T \rightarrow q_{2} T\,, 
Z \rightarrow q_{1}q_{3} Z\,.
}
The formula for the character reads 
\eqn\coE{
Z_{l=0}(q_{1},q_{2},q_{3}) = { (1 - q_{1} q_{2} q_{3}) \over (1 - q_{1}) (1 - q_{2}) (1 - q_{1} q_{2}) ( 1 - q_{1} q_{3})}\,.
 }
To compare with $Z_{l=0}(t)$ we set $q_{1} = t, q_{2}=t, q_{3} =1$. 

To compute higher orders, we can proceed by 
counting the states as above. This provides an important check 
on the formula \acD. The result for the first massive states for $N=4$ is 
$$
Z_{l=1}(q_{1}=q_{2}=t, q_{3}=1) = 7 + 24\,t + 54\,t^2 + 96\,t^3 + 150\,t^4 + 216\,t^5 + {\cal O}(t^{6})\,, 
$$
$$
Z_{l=2}(q_{1}=q_{2}=t, q_{3}=1) = 
34 + 112\,t + 243\,t^2 + 432\,t^3 + 675\,t^4 + 972\,t^5 + {\cal O}(t^{6})\,, 
$$
$$
Z_{l=3}(q_{1}=q_{2}=t, q_{3}=1) = 
132 + 416\,t + 891\,t^2 + 1568\,t^3 + 2450\,t^4 + 
  3528\,t^5 + {\cal O}(t^{6})\,.
$$
The formula \acD\ can provide the complete expression also 
for generic charges $q_{i}$, but the formula is not very illuminating and 
we do not report it here. 

\subsec{The supercone}

In the last example, we introduce some fermions. We consider the 
cone $\l \bar \l =0$ analyzed before and we add two fermions $\psi$ and $\bar\psi$. This allows us to consider two situations: {\it i)} we 
maintain the constraint $\Phi(\l)$ as in \conoA\ and we weight the 
fermions with ghost charge $+1$ (and $-1$ for their conjugates $\eta$ and $\bar \eta$). {\it ii)} we 
modifying the constraint by adding a fermionic part to it as follows
\eqn\superA{
\l \bar\lambda + \psi \bar \psi = 0\,. 
}
This constraint can be viewed as a supersymmetric version 
of the cone (or of the conifold if we add additional variables) 
\SchwarzAK\ and a higher-dimensional analogue of it 
appears in the recent formulation of N=4 SYM 
in terms of supertwistors \WittenNN\ (more precisely, the superambitwistor space is described by a superquadric defined locally by an equation of this form). 
The correct way to interpret \superA\ is in 
the context of supervariety, but for the present purposes we do not 
need any rigorous definition of \superA. 
In the first case we distinguish the variables $\lambda,\bar \lambda$ entering the constraints 
from the $\psi,\bar\psi$ by assigning them two different parameters, respectively $t$ and $s$. 
 At the end, we take the limit $s \rightarrow t$ 
in order to count the states with their statistic. 
The partition function for the first case
is given 
\eqn\superB{
Z^{(i)}(q|t) = \lim_{s\rightarrow t}
\Big[{(1 - s^{2}) (1 - t)^{2} \over (1- s)^{2} }  \times
}
$$
\prod_{n=1}^{\infty} 
{
(1 - q^{n} t)^{2} (1 - q^{n} t^{-1})^{2} 
(1 - q^{n} s)^{2}) (1 - {q^{n} s^{-2}}) 
\over 
(1 - q^{n} s)^{2} (1 - {q^{n}  s^{-1}})^{2} 
}\Big|_{\rm regular}\Big]
$$
while in the second case (when we assign the same 
number to $\psi, \bar \psi$ and to $\l, \bar\l$), 
the partition function becomes 
\eqn\superB{
Z^{(ii)}(q|t) = \lim_{s\rightarrow t}
\Big[{(1 - s^{2}) } 
\prod_{n=1}^{\infty} 
{(1 - q^{n} s^{2}) (1 - {q^{n} s^{-2}}) 
}\Big|_{\rm regular}\Big]\,.
}
 In the second case, taking the regular part of 
 the expression and the limit $s\rightarrow t$ is harmless. 
 
It is interesting to consider the partition functions at the first level. 
They are respectively 
\eqn\superC{
Z^{(i)}_{l=1}(t)  = 2 - {2 \over t} - t^2 + t^4\,, ~~~~~
Z^{(ii)}_{l=1}(t)  = 1 - t^2 + t^4\,. 
}
They have a natural interpretation in terms 
of states. We notice for example that the first two terms 
of  $Z^{(i)}_{l=1}(t)$ correspond to the currents $J = :w \l:$ and 
$\bar J = :\bar w\bar \l:$  and to the $\eta$ and $\bar \eta$ (the conjugate fermionic momenta), respectively.  
On the other side, the first term of $Z^{(ii)}_{l=1}(t)$ is related to gauge invariant 
currents. The form of these currents is due to the form of the constraints 
$\Phi(\l)$ and 
in the supercone case \superA\ 
there are 5 bosonic currents and 4 fermionic ones 
$$
J =: w \l + \eta \psi:\,, ~~~ \bar J =: \bar w \bar\l + \bar\eta \bar\psi:\,, ~~~
N = :w \l - \bar w \bar \l:\,, ~~~~
K = \eta \bar \psi\,, ~~~~ \bar K = \bar \eta \psi\,,
$$
$$
\Omega = w \psi - \bar \eta \bar \l\,, ~~~~
\bar \Omega = \bar w \bar \psi - \eta \l\,, ~~~
\Xi = w \bar \psi + \eta \bar \l\,, ~~~~
\bar \Xi = \bar w \psi + \bar \eta \l\,. 
$$
The remaning terms of the two partition functions agree. They are the contributions of the  
states formed only with $\l, \bar\l, \psi, \bar \psi$ and 
there is one-to-one correspondence between the two sets of states in the two cases. 
We also notice that in the first case with have negative powers 
of $t$ since there is no restriction on the powers of $\eta, \bar\eta$ 
and their derivatives. This shows once again that removing the poles in $s$ is not the same as removing the poles in $t$. 

\newsec{Conclusions and Outlook}

We presented here  a formula to compute the partition function 
of a chiral sigma model on a constrained surface. We limit ourselves to a 
specific form of constraint: irreducible, quadratic and holomorphic. Even 
with these restrictions, there are interesting models to analyze. One 
of the main motivations of this analysis is the construction of the 
partition function using the Koszul theorem to resolve the constraints and 
to apply the technique to pure spinor string theory. Already for the set of 
constraints we consider the implementation of the Koszul theorem at the 
level of partition function is non-trivial. We used these examples to test our formula, since we 
can compute the spectrum independently with a different method. 
The next step will be to analyze the case of reducible constraints and 
to specialize it to the case of pure spinors for string theories. We leave it to future 
work to prove that our prescription gives the correct result also in the most general situation. 

There are also other applications that it would be interesting to analyze, such as 
constraints of higher degree and/or submanifolds of higher codimension. 
One feature that has emerged, in particular from the analysis of the supercone, 
is the following: at the level of zero modes the character formula is only sensitive to the degree 
of the polynomial defining the constraint, and not to the detailed form of the polynomial. 
In fact, the Koszul method introduces a single ghost for implementing the constraint, and the ghost only 
remembers the degree via the charge assignment. This is reminiscent of the dependence of the Landau-Ginzburg models on the superpotential.  
However, the details of the polynomial $\Phi(\l)$ are seen at the first massive level. Indeed, we have showed that the number of independent gauge invariant operators is related to the precise form of the constraint. It remains to be understood whether this feature is specific to the case of supervarieties. 

\bigskip
\noindent{\bf Acknowledgments}
We thank F. Morales, Y. Oz and E. Scheidegger 
for useful discussions. G. P. thanks 
Universit\`a del Piemonte Orientale at Alessandria where 
this work was completed.  
This work is partially supported by the European Community's Human Potential
Programme under contract MRTN-CT-2004-005104 (in which P.A.G. is
associated to Torino University) and by the Italian MIUR under contract PRIN-2005023102. G.P. is supported by the SFB 375 of DFG.

\listrefs
\bye